\documentclass[aip,amsmath,amssymb,reprint]{revtex4-1}

\usepackage{graphicx}% Include figure files
\usepackage{dcolumn}% Align table columns on decimal point
\usepackage{bm}% bold math
\usepackage[utf8]{inputenc}
\usepackage[T1]{fontenc}
\usepackage{mathptmx}
\usepackage{etoolbox}

\makeatletter
\def\@email#1#2{%
 \endgroup
 \patchcmd{\titleblock@produce}
  {\frontmatter@RRAPformat}
  {\frontmatter@RRAPformat{\produce@RRAP{*#1\href{mailto:#2}{#2}}}\frontmatter@RRAPformat}
  {}{}
}%
\makeatother
\begin{document}

\preprint{AIP/123-QED}

\title[]{RLC resonator with diode nonlinearity: Bifurcation comparison of numerical predictions and circuit measurements}
% Force line breaks with \\
\author{Edward H. Hellen}
 \email{ehhellen@uncg.edu}
% \altaffiliation[Also at ]{}%Lines break automatically or can be forced with \\
%\author{B. Author}%
% \email{Second.Author@institution.edu.}
\affiliation{Department of Physics \& Astronomy, University of North Carolina Greensboro, Greensboro, NC USA%\\This line break forced with \textbackslash\textbackslash
}%

%\author{C. Author}
% \homepage{http://www.Second.institution.edu/~Charlie.Author.}
%\affiliation{%
%Second institution and/or address%\\This line break forced% with \\
%}%

\date{\today}% It is always \today, today,
             %  but any date may be explicitly specified

\begin{abstract}
A nonlinear RLC resonator is investigated experimentally and numerically using bifurcation analysis. The nonlinearity is due to the parallel combination of a semiconductor rectifier diode and a fixed capacitor. The diode's junction capacitance, diffusion capacitance, and DC current-voltage relation each contribute to the nonlinearity. The closely related RL-diode resonator has been of interest for many years since its demonstration of period-doubling cascades to chaos. In this study a direct comparison is made of dynamical regime maps produced from simulations and circuit measurements. The maps show the variety of limit cycles, their bifurcations, and regions of chaos over the 2-d parameter space of the source voltage's frequency and amplitude. The similar structures of the simulated and experimental maps suggests that the diode models commonly used in circuit simulators (e.g., SPICE) work well in bifurcation analyses, successfully predicting complex and chaotic dynamics detected in the circuit. These results may be useful for applications of varactor-loaded split ring resonators. 
\end{abstract}

\maketitle

\begin{quotation}
Behavior of the resistor-inductor-diode oscillator is a popular example of complexity and chaos. The nonlinearity of the semi-conductor junction is accounted for by well-known models used regularly in circuit simulators such as SPICE.\cite{rect-appl-hb} This nonlinear oscillator is also the basis for nonlinear split ring resonators, which are of interest due to interesting magnetic properties produced when used to form arrays in metamaterials. Here, bifurcation analysis and Lyapunov exponents are used to calculate a dynamical regime map in a 2-d parameter space of the source frequency and amplitude which is compared to a regime map created from circuit measurements. Good agreement is found on the structure of the maps as indicated by the types of limit cycles, their bifurcations, and period-doubling cascades to chaos. The results suggest that these types of bifurcation analyses may be useful for developing applications of the nonlinear split ring resonators in metamaterials. 
\end{quotation}


\section{Introduction}
The series RLC circuit is one of the most common examples of the damped harmonic oscillator. In this numerical and experimental study, nonlinearity is introduced into the resonator by including a diode in parallel with the capacitance. Semiconductor diodes have nonlinear current-voltage relationships which are accounted for by mathematical models used in the well-known SPICE modeling tool for circuit design.\cite{rect-appl-hb} The nonlinearities cause the circuit oscillations to have complex responses to a sinusoidal driving voltage. In this work, bifurcation analysis and Lyapunov exponents are used to find the limit cycles and chaotic attractors of this RLC-diode circuit and their dependence on the frequency and amplitude of the driving source. These simulation results are then compared to circuit measurements. The goal here is to determine how well the bifurcation and Lyapunov exponent analysis can predict actual circuit behavior over a wide range of parameter space. 

The circuit considered here is essentially the same as the resistor-inductor-diode (RLD) circuit used earlier to demonstrate experimentally the period-doubling cascades to chaos which can be induced by varying parameters of the sinusoidal voltage source.\cite{Linsay1981,Testa1982} These experiments also demonstrated the spacing convergence of the period doubling bifurcation parameter values predicted by Feigenbaum's theory on the universal behaviour of period doubling systems.\cite{Feigenbaum1978,Feigenbaum1979} Computer calculations based on the RLD circuit analysis also predicted the period doubling cascade to chaos.\cite{Azzouz1983} Intermittency routes to chaos were demonstrated.\cite{Jeffries1982} Comparison of model calculations and experimental measurements were made using bifurcation diagrams, return maps, and Poincar\'{e} plots.  \cite{Brorson1983,Klinker1984} A simplified diode model was used to calculate a 2-dim bifurcation regime map which agreed well with experimentally observed bifurcations.\cite{TANAKA1987} 

These early studies' good agreement of theory and experiment for this simple circuit's complex behavior resulted in the RLD circuit becoming a popular example of a chaotic system. A good example relevant to this paper is the comparison of numerical predictions and circuit measurements by Carroll and Pecora showing good agreement on the location of the $1^{st}$ period-doubling in 2-dim parameter space.\citep{Carroll2002} 

In the years since the above studies, bifurcation analysis tools have improved and computing power has increased.   It is somewhat surprising that more use has not been made of these tools to make detailed predictions of the dynamical regimes of the RLD circuit by using the diode's well established mathematical models used in SPICE. To the author's knowledge, a direct comparison of numerically calculated and experimentally measured bifurcation regime maps, delineating the various limit cycles over a 2-dim parameter space has not been done for the RLD circuit. 

In the past couple of decades, there has been increasing interest in exploiting nonliner RLC circuits in metasurfaces for their interesting magnetic properties.\cite{Pendry1999} In these applications, the RLC circuit is a split ring resonator (SRR) with nonlinearity introduced into the capacitance, often as a semiconductor junction, \cite{shadrivov2006,wang2008,lazarides2011,rose2011} thereby making the SSR a RLD. The SRRs are typically small (dimensions of $\lessapprox$ 10s of $\mu m$) and are driven by the electromotive force induced by electromagnetic waves in the GHz or THz range. The system considered here is conceptually and mathematically the same as a SRR. Only the size and frequency differs. Here, a function generator drives a circuit constructed from common electronic components: a resistor, an inductor, and the parallel combination of a fixed capacitor and a 1N4003 rectifier diode. The diode's semiconductor junction is the source of the nonlinearity, just as it is for those SRRs which incorporate varactor diodes.

The early studies of the RLD were primarily motivated by scholarly interest in understanding the nature of their complex and chaotic behavior. The more recent studies of arrays of the nonlinear SRRs used in metasurfaces are aimed at applications and therefore may benefit from more accurate modeling of the nonlinear SRR. 

The goal of this work is to compare simulations and circuit measurements of the RLC-diode over the 2-dim parameter space of the source's amplitude and frequency. The simulations use bifurcation tools and Lyapunov exponent calculations to discover the limit cycles, their stability, and chaotic regions over the 2-dim parameter space. The model used for the diode accounts for junction capacitance, diffusion capacitance, and the nonlinear current-voltage relation by using well-established models used in SPICE. Circuit measurements detecting period-doubling, hysteresis, and chaotic-like behavior are made using standard digital function generator and oscilloscope. The work here refers to the RLC-diode circuit because the diode is in parallel with a fixed capacitor whereas the RLD circuit generally does not include a fixed capacitor. The fixed capacitor allows for additional flexibility in lowering the resonance frequency of the circuit but does not change the essential nature of comparing the predictions and measurements of complex behaviors caused by the presence of the diode. 

The presentation style here attempts to be accessible to those who are interested in the RLD circuit but are not particularly familiar with bifurcation analysis methods and terminology. Therefore descriptions of the bifurcation methods are restricted to relevance to this work and may lack a more general mathematical completeness and rigor. 
\section{RLC-diode Equations}
\begin{figure} [h]
\includegraphics[width=3.2 in]{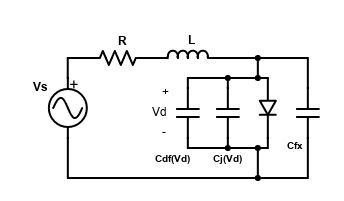}
\caption{Driven nonlinear RLC-diode circuit where a diode provides the nonlinearity. Diode is modeled by parallel combination of junction capacitance, diffusion capacitance, and DC diode (the diode symbol). A fixed capacitance in parallel with the diode is also included. R represents the sum of explicit resistance and the inductor's intrinsic resistance. }
\label{circuit}  
\end{figure}

The system of interest shown in Fig.\ \ref{circuit} is a sinusoidally driven series RLC-diode circuit with a voltage-dependent capacitance due to the diode. We investigate how the dynamics of the circuit oscillations depend on the amplitude and frequency of the driving source Vs. The focus is on using bifurcation analysis methods to identify the limit cycles and determine their stability.

The capacitance of the RLC-diode resonator is the result of the parallel combination of the diode and the fixed capacitance (Cfx in the figure) which includes stray capacitance. The mathematical model used for the diode is the parallel combination of junction capacitance Cj, diffusion capacitance Cdf, and the diode's standard DC current-voltage relation. All three of these terms are voltage dependent and therefore contribute to the nonlinearity of the system. The well-known models of semiconductor diodes used in the SPICE circuit simulator are used to account for these three effects. 

For a voltage source $V_s=A\cos (\Omega t)$ driving the RLC-diode resonator, summing voltages around the circuit gives
\begin{equation}
V_s=L\frac{dI}{dt}+IR+V_d
\label{V-loop}
\end{equation}
where $V_d$ is the voltage across the parallel combination of the diode and fixed capacitor. 

Current $I$ from the voltage source splits, supplying current $dQ/dt$ to the junction capacitance, current $I_{df}$ to the diffusion capacitance, current $I_{fx}$ to the fixed capacitor, and current $I_d$ to the DC diode, giving the equation 
\begin{equation}
I=\frac{dQ}{dt}+\frac{dQ_{df}}{dt}+\frac{dQ_{fx}}{dt}+I_d.
\label{i-node}
\end{equation}
Source current $I$ and junction capacitance charge $Q$ are the two variables chosen to describe the system. Therefore, voltages and other charges must be expressed in terms of $I$ and $Q$. 

$I_{df}$ and $I_{fx}$ are expressed in terms of $Q$ by using 
\begin{equation*}
\frac{dQ_i}{dt}=\frac{dQ_i}{dV_d}\frac{dV_d}{dt}=C_i\frac{dV_d}{dt}=C_i\frac{dV_d}{dQ}\frac{dQ}{dt}
\end{equation*}
where incremental capacitance $C_i=dQ_i/dV_d$ is used and subscript $i=(df,fx)$ for the diffusion and fixed capacitances. Solving Eq.\ \eqref{i-node} for the current into the junction capacitance gives
\begin{equation}
\frac{dQ}{dt}=\frac{I-I_d}{1+(C_{df}(V_d)+C_{fx})\frac{dV_d}{dQ}}.
\label{Q-eqn}
\end{equation}
The Appendix shows that the SPICE diode model allows $V_d$ and $I_d$ to be expressed in terms of $Q$. Thus, Eqs.\ \eqref{V-loop} and \eqref{Q-eqn} are the foundation for modeling the circuit in Fig.\ \ref{circuit} in terms of the source current $I$ and the diode junction capacitance charge $Q$. 

Equations \eqref{V-loop} and \eqref{Q-eqn} are converted to a dimensionless form by defining dimensionless variables,  
\begin{equation}
q\equiv\frac{Q}{Q_0},\: i\equiv\frac{I}{V_J}\sqrt{\frac{L}{C_{J0}+C_{fx}}},\: v\equiv\frac{V}{V_J},\: t'\equiv\Omega_0 t
\label{dimensionless}
\end{equation}
where $V_J$ and $C_{J0}$ are junction capacitance SPICE parameters (see Appendix). $Q_0=V_JC_{J0}$ is a naturally defined junction charge parameter and $\Omega_0=1/\sqrt{L(C_{J0}+C_{fx})}$ is the nominal resonant frequency. Renaming $t'$ as $t$ results in
\begin{subequations}
\begin{align} 
\frac{dq}{dt}= &\: \left(i-i_d(q)\right)\frac{C_{J0}+C_{fx}}{C_{J0}+(C_{df}(q)+C_{fx})\frac{dv_d(q)}{dq}}\\
\frac{di}{dt}= & -v_d(q)-\sigma i+\mu \cos (\omega t)
\end{align}
\label{nonauton-eqns}
\end{subequations}
where $t$ is now dimensionless. The dependence of $v_d$ on the variables $(q,i)$ is required in Eq.\ \eqref{nonauton-eqns}. The Appendix shows that:
\begin{equation}
v_d(q)=
\begin{cases}
1-\left(1-(1-M)q\right)^{\frac{1}{1-M}}\; & \text{for }q < q_t\\
FC+(1-FC)^M(q-q_t) & \text{for }q> q_t
\end{cases}
\label{vd(q)}
\end{equation}
and
\begin{equation}
\frac{dv_{cap}(q)}{dq}=
\begin{cases}
\left(1-(1-M)q\right)^{\frac{M}{1-M}}\; & \text{for }q < q_t\\
(1-FC)^M & \text{for }q> q_t
\end{cases}
\end{equation}
where 
\begin{equation}
q_t=\frac{1-(1-FC)^{(1-M)}}{1-M}.
\end{equation}
$M$ and $FC$ are two more junction capacitance SPICE parameters. The case for $M=1$ is in the Appendix.

The diode's DC current-voltage relation requires two SPICE parameters, reverse saturation current $I_S$ and emission coefficient $N$. The dimensionless relation is
\begin{equation}
i_d(q)=i_S\left(e^{40V_jv_d(q)/N}-1\right)
\end{equation}
where Eq.\ \eqref{dimensionless} is used to obtain the dimensionless $i_S$ from $I_S$.
The diffusion capacitance is
\begin{equation}
C_{df}(q)=\frac{40TT\times I_S}{N}e^{40V_Jv_d(q)/N}.
\end{equation}
where $TT$ is the SPICE parameter for forward transit time. 

We use SPICE parameter values appropriate for the 1N4001-1N4004 family of rectifier diodes: $I_S=70$ pA, $N=1.4$, $C_{J0}=33$ pf, $V_J=0.35$ volts, $M=0.45$, $FC=0.5$, and $TT=5\mu s$.   

Remaining parameters in Eq.\ \eqref{nonauton-eqns} are
\begin{equation}
\sigma=R\sqrt{\frac{C_{J0}+C_{fx}}{L}},\; \mu=\frac{A}{V_J},\; \omega=\frac{\Omega}{\Omega_0}.
\label{sig-mu-omega}
\end{equation}
In this work $\omega$ and $\mu$ are the parameters of interest to vary in the bifurcation analysis. Equation \eqref{nonauton-eqns} is the non-autonomous version of the equations for our driven RLC-diode circuit.  

Some of the bifurcation analysis methods are more easily suited to autonomous systems. Therefore, Eq.\ \eqref{nonauton-eqns} is converted to an autonomous form by including two additional variables $u(t)$ and $v(t)$. The differential equations for these variables are constructed so that $u$ and $v$ are sinusoidal with frequency $\omega$ and unity amplitude.\cite{ermentrout} The autonomous version of system \eqref{nonauton-eqns} is then
\begin{subequations}
\begin{align} 
\frac{dq}{dt}= &\: \left(i-i_d(q)\right)\frac{C_{J0}+C_{fx}}{C_{J0}+(C_{df}(q)+C_{fx})\frac{dv_d(q)}{dq}}\\
\frac{di}{dt}= & -v_d(q)-\sigma i+\mu u\\
\frac{du}{dt}= & u(1-u^2-v^2)-\omega v\\
\frac{dv}{dt}= &\: v(1-u^2-v^2)+\omega u.
\end{align}
\label{auton-eqns}
\end{subequations}
Initial conditions should satisfy $u^2+v^2=1$. The sinusoidal behavior of $u$ and $v$ allows for periodic behavior of all 4 variables. This addition of two variables is in contrast to converting to an autonomous system by adding a single variable $u(t)=t$ which results in one fewer equation, but has the disadvantage of not allowing periodic behavior of all the variables. 

The existence of the sinusoidal driving term insures there will be no fixed points (except for the trivial case of $\mu$ or $\omega=0$) and that the main limit cycle is period-1, meaning an oscillation with the same period as the sinusoidal source.

\section{Analysis Method}
Bifurcation analysis of Eq.\ \ref{auton-eqns} was done using XPPAUT \cite{ermentrout} and AUTO-07p. \cite{auto-07p} These software tools are used to find the limit cycles (LC), their bifurcation points, and 2-dim parameter space regime maps showing the types of LCs. Time-series integration was done using an adaptive 4\textsuperscript{th}-order Runge-Kutta routine. Lyapunov exponents were calculated using the method of Wolf et. al. \cite{wolf1985} The bifurcation analysis consists of generating bifurcation continuation plots and the 2-dim regime maps. The bifurcations for Eq.\ \ref{auton-eqns} are period-doublings (PD) and limit points (LP). Here we describe the bifurcation points and their relation to the bifurcation continuation plots. It is worth noting that while the diode's mathematical model used here in the above software tools uses SPICE parameters, the SPICE software itself was not used for any of the numerical simulations. 

PD points occur on a bifurcation continuation plot at parameter values where a stable LC becomes unstable and a LC with double the period is created. The main LC of Eq.\ \ref{auton-eqns} is period-1, having the same period as the source. This LC exists as either stable or unstable over the entire parameter space. Figure 
\begin{figure} [h]
\includegraphics[width=3.2 in]{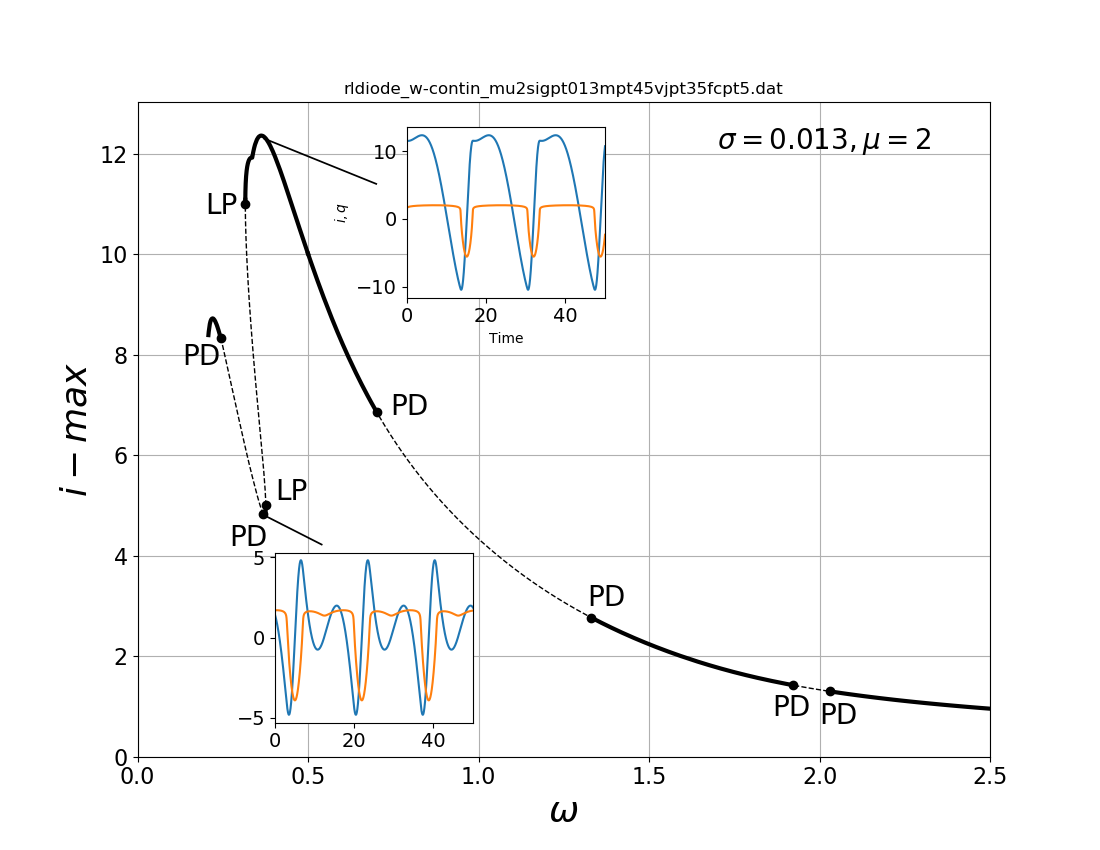}
\caption{Numerical $\omega$-continuation bifurcation plot showing the maximum value of $i$ for the LC. $\mu=2$ and $\sigma=0.013$. Solid (dashed) line indicates stable (unstable) LC. PD and LP bifurcation points are indicated. Insets show coexisting LC time-series of $q$ (orange) and $i$ (blue) at $\omega=0.372$.}
\label{w-contin-sig.01}  
\end{figure}
\ref{w-contin-sig.01} shows the $\omega$-continuation bifurcation plot for the main LC. Six PD points are seen, where the LC changes stability. Solid (dashed) line indicates stable (unstable) LC. The plot shows how $i$-max (the maximum value of variable $i$'s oscillation) depends on parameter $\omega$. LPs (also called saddle-node bifurcations) occur on a bifurcation continuation plot at parameter values where a stable LC branch and an unstable LC branch extending together in parameter space reach an extreme parameter value (the LP) where they collide and disappear. Another description is that a LC branch reverses its direction along the parameter-axis at the LP, with concomitant change in stability. 

The two LPs in Fig.\ \ref{w-contin-sig.01} create a region of hysteresis between their locations at $\omega=0.316$ and 0.376, where there is coexistence of large and small amplitude LCs as demonstrated by the time-series insets. The large amplitude LC is stable over the hysteresis region, while the small amplitude LC becomes unstable for $\omega$ below 0.369 where it undergoes period-doubling. 

The LP and PD bifurcation points detected in the parameter continuation plots are the starting points for tracing their paths in the 2-dim $\omega$-$\mu$ parameter space. The paths are the borders for different dynamical regimes. Creation of these dynamical regime maps is a goal of this study because of their usefulness in presenting how the system Eq.\ \ref{auton-eqns} behaves throughout the $(\omega,\mu)$ parameter space. 

$\omega$-continuation bifurcation plots of stable period-doubled LCs are used to find higher order period-doubling points. Those points are then traced in the 2-dim  $\omega$-$\mu$ parameter space to find their borders allowing determination of period-doubling cascade regions. 

The Lyapunov exponents (LE) are useful for detecting regions with no stable LCs, where the stable oscillatory states are either quasi-periodic or chaotic. Borders for chaotic regions are not detected by the bifurcations tools but are easily identified in LE plots as parameter values where the maximal LE changes from zero (indicating stable LC) to positive (indicating chaos).     

\section{Circuit}\label{circuit section}
Circuits were constructed based on Fig.\ \ref{circuit}. The inductor was $L=10$ mH with $23\Omega$ of intrinsic resistance. The diode was 1N4003, a standard rectifier. Fifteen of these diodes were used and in-house measurements were made of the zero-bias junction capacitance, the reverse breakdown voltage, and the reverse recovery time. Junction capacitance ranged from 23 to 85 pf, breakdown voltage was 300 to 1400 volts, and reverse recovery time was 8 to 32 $\mu s$. All 15 diodes were were tested in RLC-diode circuits to check the consistency of the dynamical regime results when using different diodes and to determine the sensitivity to the diode properties. 
%\begin{figure} [h]
%\includegraphics[width=3.2 in]{rlc-circuit.png}
%\caption{One of the RLC-diode circuits used for measurements. Resistor is $100\Omega$, L =10 mH, and 1N4003 diode in parallel with 68 pf. Inductor has $23\Omega$ making $R=123\Omega$. Measured resonance, $f_0=140kHz$.}
%\label{circuit-pic}  
%\end{figure}

The capacitor in parallel with the diode was 68 pf. The fixed capacitance $C_{fx}$ in Fig.\ \ref{circuit} accounts for both the 68 pf capacitor and stray capacitance. Stray capacitance was determined by measuring the resonant frequency for a small signal diode with a nominal 1 pf junction capacitance, yielding stray capacitance of 24 pf. The resistor was $100\Omega$ making the total series resistance $R=123\Omega$. Using $123\Omega$ with $C_{fx}=92pf$ and $C_{J0}=33pf$ (for the particular diode used to create the experimental results regime map) in Eq.\ \ref{sig-mu-omega} gives $\sigma=0.013$ for use in simulations.

\begin{figure} [h]
\includegraphics[width=3.2 in]{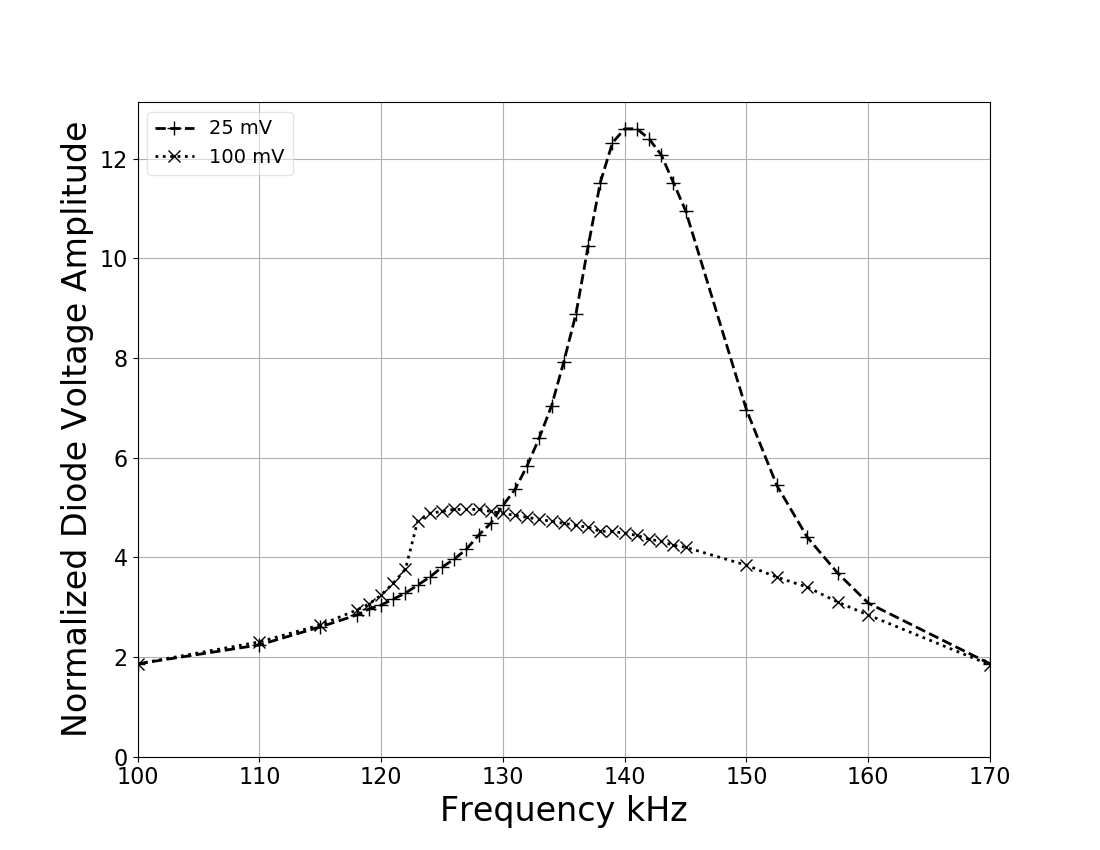}
\caption{RLC-diode circuit's frequency response. Normalized diode amplitude for 25 and 100 mV source amplitudes. $R=123\Omega$, L =10 mH, and 1N4003 diode in parallel with 68 pf capacitor and 24 pf stray capacitance.}
\label{circuit-resonance}  
\end{figure}

Resonance frequency was determined using a small amplitude (25 mV) source. Figure \ref{circuit-resonance} shows the frequency response of the diode voltage amplitude for source amplitudes of 25 and 100 mV.  
The voltage amplitude has been normalized by the source amplitude. The 25 mV source plot shows a typical resonance response, with a resonant frequency of 140 kHz. The maximum forward bias is about $12\times 25mV=0.3$ volts. The 100 mV source plot shows the distorted resonance response which is due to the significantly higher diode current at the maximum forward bias of $5\times 100mV=0.5$ volts.   

\section{Results}
\subsection{Numerical Bifurcation Analysis}
\begin{figure} [h]
\includegraphics[width=3.2 in]{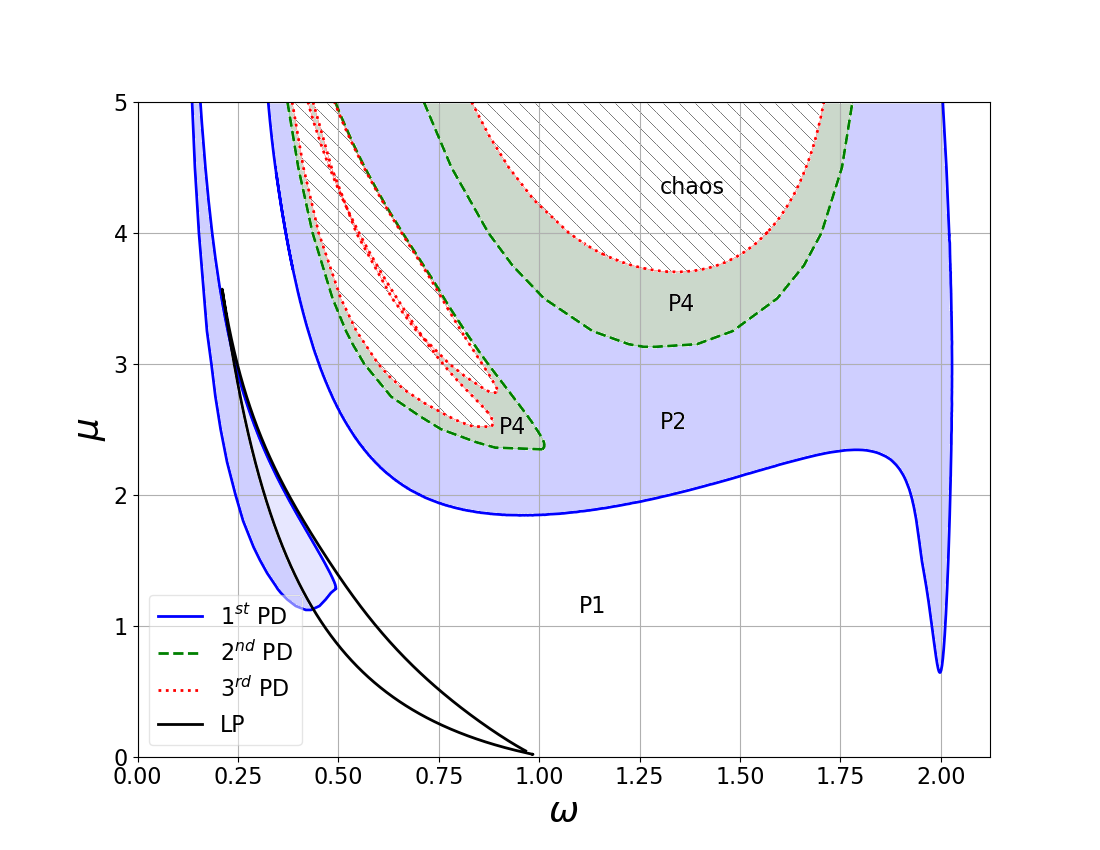}
\caption{Numerical $\omega$-$\mu$ regime map for $\sigma=0.013$, $C_{fx}=92$ pf. P1 (period-1), P2, and P4 LC regions are indicated. Hatching indicates higher order doublings and chaos. Frequency $\omega$ and source amplitude $\mu$ are dimensionless.}
\label{sigma-0.01-regime map}  
\end{figure}

Figure \ref{sigma-0.01-regime map} is the numerical dynamical regime map showing the limit cycles and chaotic oscillations of the RLC-diode circuit in the 2-dim $\omega$-$\mu$ parameter space. The map consists of LP and PD lines. Traversing the map at $\mu=2$, there are six crossings of $1^{st}$ PD lines and two crossings of LP lines, corresponding to the six PD points and two LP points in the $\omega$-continuation of Fig.\ \ref{w-contin-sig.01}. 

A region of hysteresis of the period-1 (P1) main LC is indicated by the LP lines. Inside this narrow region, two versions of the main LC coexist, which causes a jump in the LC amplitude during a sweep of the frequency when a LP is encountered causing a sudden switch from one version to the other. Figure \ref{w-contin-sig.01} shows the hysteresis and the time-series of the two coexisting P1 LCs for $\mu=2$.  

There are two regions of period doubling of the P1 LC, a narrow region at low frequency and a much larger region which occupies a substantial amount of the map. The larger PD region contains two regions of higher order period doubling cascades. The $2^{nd}$ and $3^{rd}$ doubling lines are shown. Period doubling cascades to chaos are common when the spacing between higher order doublings decreases, as is the case here. The $3^{rd}$ period doubling line can often be used as an approximation for the doubling cascade's onset of chaos. Confirmation of chaos by calculation of Lyapunov exponents is shown below. 

\subsection{Lyapunov Exponents}
\begin{figure} [h]
\includegraphics[width=3.2 in]{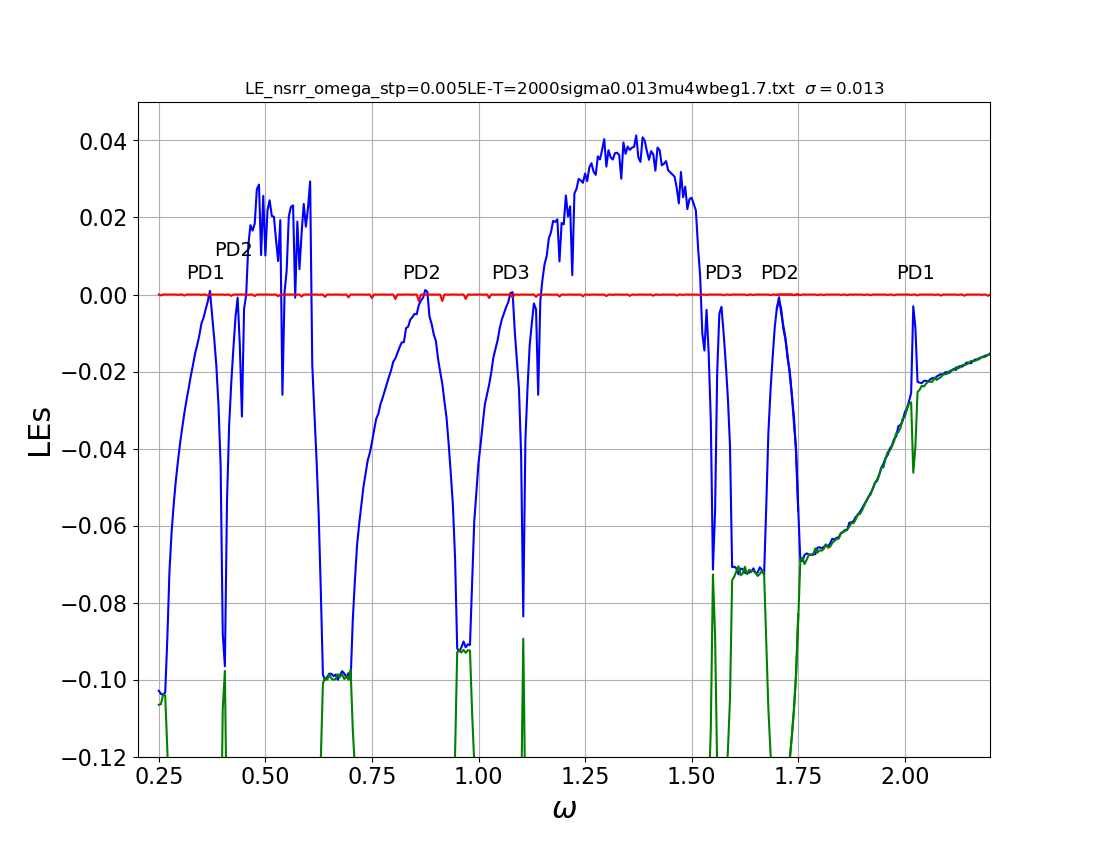}
\caption{Three maximal LEs for increasing source frequency $\omega$ with $\sigma=0.013,\mu=4$. Chaos occurs where the blue LE is positive. $1^{st}$, $2^{nd}$, and $3^{rd}$ period-doubling points are indicated.}
\label{LE}  
\end{figure}

LE calculations detect parameter ranges with chaotic behavior and are therefore complementary to the parameter regime maps created by bifurcation analysis since those methods do not detect chaos. In LE plots for an autonomous system, stable LC is indicated by the maximal LE being zero, whereas chaotic behavior has a positive maximal LE. The LE plot for source amplitude $\mu=4$ in Fig.\  \ref{LE} shows there are two regions where the maximal LE is positive (where the blue LE exceeds the red LE), near $\omega=0.5$ and a wider region around $\omega=1.35$. These two regions confirm the period doubling cascades to chaos suggested by the nested higher order doublings in Fig.\ \ref{sigma-0.01-regime map}. 

Information about PD points can also be indicated on the LE plot. When a changing parameter causes a stable LC to undergo period doubling, the maximal LE remains at zero throughout the process. However, the second maximal LE increases until it reaches zero at the PD bifurcation point, then it returns to increasingly negative numbers. 
Many such occurrences are labeled in Fig.\ \ref{LE} based on their agreement with where PD lines are crossed at $\mu=4$ in Fig.\ \ref{sigma-0.01-regime map}.  

\begin{figure} [h]
\includegraphics[width=2 in]{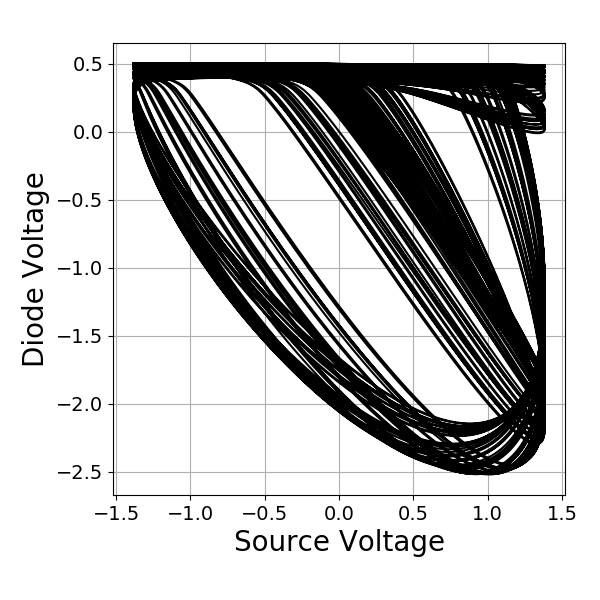}
\caption{Numerical phase plot for $\mu=3.95,\omega=1.25,\sigma=0.013$, located inside the period-doubling cascade.}
\label{sim-phase-plot}  
\end{figure}

Figure \ref{sim-phase-plot} shows the simulated phase plot of source voltage versus diode voltage at $\omega=1.25,\mu=3.95$ inside the right-side period doubling cascade to chaos in Fig.\ \ref{sigma-0.01-regime map}. The phase plot shows the ``filling in" of regions of the phase space characteristic of chaos. 

The LE plot gives information only for the dynamical state whose basin of attraction has captured the system. This is in contrast to the regime maps which can show bifurcation lines for coexisting states, regardless of their basins of attraction. This means that bifurcation lines on the regime maps might not be indicated in a LE plot. An example is the right side $2^{nd}$ PD of the left doubling cascade in Fig.\ \ref{sigma-0.01-regime map} which shows this PD at $\omega=0.66$ for $\mu=4$. The $3^{rd}$ PD is very close to the $2^{nd}$, suggesting the cascade to chaos should also occur close to $\omega=0.66$. However, Fig.\ \ref{LE} shows that chaos extends to only 0.61. The existence of the two $3^{rd}$ PD lines which overlap indicates a complicated coexistence of higher order doublings and two separate cascades to chaos. Presumably, the basin of attraction of a stable period doubled LC is able to capture the chaotic state at $\omega=0.61$. 

\subsection{Circuit Measurements}
\begin{figure} [h]
\includegraphics[width=3.2 in]{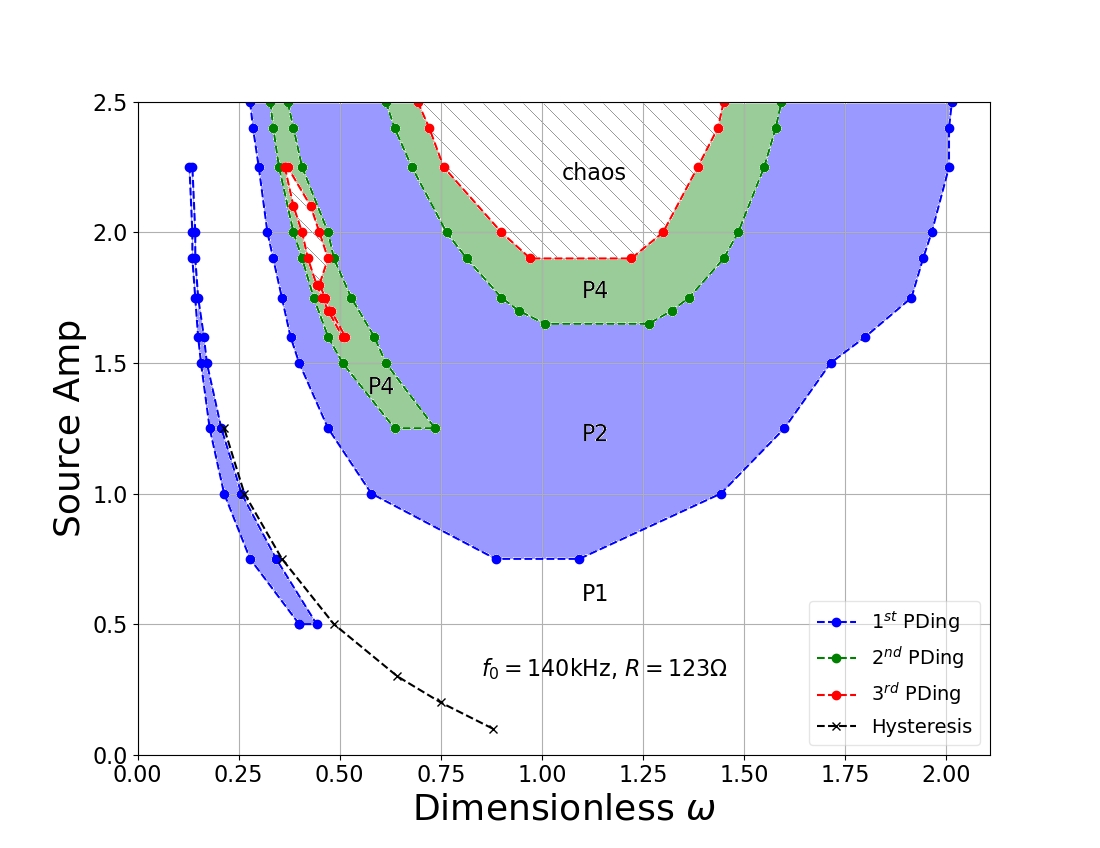}
\caption{Experimental $\omega$-``source amplitude" regime map showing PD borders and hysteresis induced amplitude jumps. Measurements from 1N4003 diode, L =10 mH, and $R=123\Omega$. $f_0=140kHz$.}
\label{1N4001-regime map}  
\end{figure}

PD points were detected as a function of source frequency and amplitude by varying source frequency at each source amplitude while monitoring the diode voltage waveform. The change in waveform indicating period doubling was easily recognized. Figure \ref{1N4001-regime map} shows the results for the circuit in Fig.\ \ref{circuit} using $R=123\Omega$ ($100\Omega$ resistor added to inductor's $23\Omega$) and 68 pf capacitor (added to 24 pf stray capacitance giving $C_{fix}=92$ pf), giving resonance frequency 140 kHz. The 10 mH inductor is known to have nonideal properties (parasitic capacitance) which become problematic for frequencies above 300 kHz.

Hysteresis was detected by noting the frequencies at which there was a large jump in the amplitude of the diode voltage LC. The location of the hysteresis for the circuit agrees well with the region in the numerical simulation, although it is much narrower for the circuit measurements. 

The overall structure of the regime maps in Figs.\ \ref{sigma-0.01-regime map} and \ref{1N4001-regime map} for simulations and circuit data have much in common. It is also informative to compare the time series of the simulations and the circuit measurements. 
\begin{figure} [h]
\includegraphics[width=3.2 in]{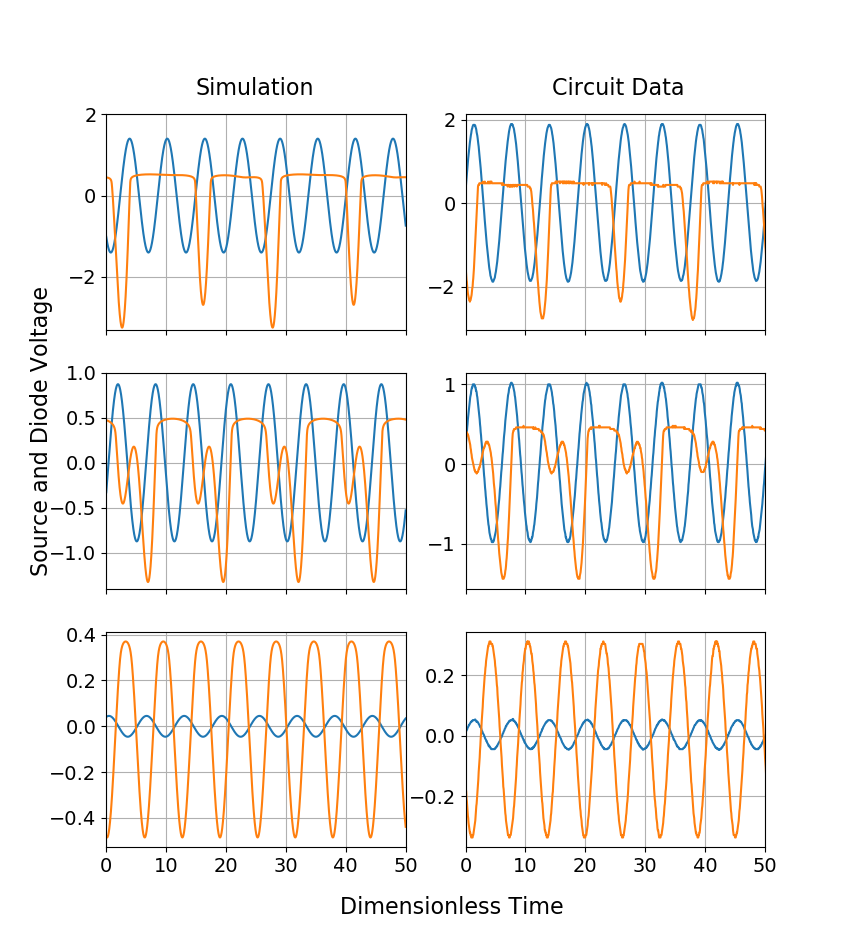}
\caption{ Time series simulations (left column) and circuit measurements (right column) for 1N4003, L =10 mH, and $R=123\Omega$ showing period-doublings as source amplitude increases for $f_0=140kHz$. Blue (orange) is source (diode) voltage. Bottom row is main LC, period-1. Middle row is period-2 LC after $1^{st}$ PD. Top row is period-4 LC after $2^{nd}$ PD.}
\label{time series compare}  
\end{figure}
Figure \ref{time series compare} shows good agreement of the LC period doublings which occur when source amplitude is increased at the nominal resonance frequency $\omega=1$ in both the numerical simulations of Fig.\ \ref{sigma-0.01-regime map} and the circuit measurements in Fig.\ \ref{1N4001-regime map}. The bottom row shows that for the small source amplitude (blue) the diode voltage LC (orange) has the same period as the source and is nearly sinusoidal. The middle row shows the time series for the period doubled P2 LC. The general shape of the doubled LC is very similar for simulation and circuit measurement, and the LC's period clearly requires 2 oscillations of the source. The top row shows the second doubling to P4 LC, which means the LC requires 4 oscillations of the source. Again, the general shapes of simulation and measurement have good agreement. 

The one large discrepancy between simulation and circuit measurement is the behavior at high frequency and low source amplitude where simulations show the P1 period doubling region extending sharply down at twice the resonant frequency.  

The existence of chaotic behavior in the circuit was investigated by continuing to increase the source amplitude beyond that used in Fig.\ \ref{time series compare} so as to complete the period doubling cascade to chaos presumed to exist inside the nested doublings. 
\begin{figure} [h]
\includegraphics[width=3.2 in]{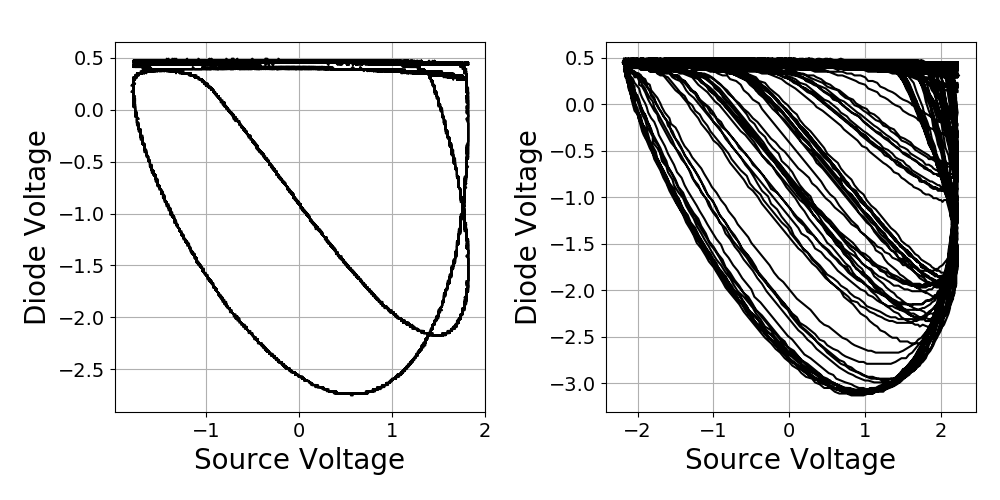}
\caption{Phase plots of measured diode voltage versus source voltage. Left is period-4 LC with 1.8-volt source amplitude, right is likely chaos with 2.2-volt source amplitude. $f=140kHz$.}
\label{circuit-phase-plots}  
\end{figure} 
At $\omega=1,(f=140kHz)$, Fig.\ \ref{1N4001-regime map} suggests that a source amplitude just above 2-volts is likely beyond the onset of chaos. The phase plots in Fig.\ \ref{circuit-phase-plots} show the behavior with 1.8 volt source, between the $2^{nd}$ and $3^{rd}$ doublings, prior to chaos (left panel), and with 2.2 volt source (right panel). The left panel shows the expected period-4 LC created by two period doublings, and the right panel shows behavior whose appearance is consistent with chaos. There is strong similarity between the right panel and the simulated chaotic phase plot in Fig.\ \ref{sim-phase-plot}.

As mentioned in Section \ref{circuit section}, fifteen 1N4003 diodes were examined which had a wide range of measured diode properties. All the diodes reproduced the main features seen in the Fig.\ \ref{sigma-0.01-regime map} numerical map, in particular, the cascade to chaos in the larger period doubling region containing $\omega=1$. However, there was some variation in the details. For example, the narrower left-side period doubling cascade in Fig.\ \ref{1N4001-regime map} located near $\omega=0.5$ shows a small island of the $3^{rd}$ PD. Some diodes only had the $2^{nd}$ doubling to period-4 LC before reverse doubling back to period-2 LC. And 3 of the 15 missed this narrow $2^{nd}$ doubling regime entirely. These 3 diodes all had relatively small reverse recovery times out of the 15 diodes. 

\section{Discussion}
Simulations and circuit measurements were compared for the driven nonlinear RLC-diode circuit where the nonlinearity is due to a rectifier diode in parallel with a fixed capacitor. Three nonlinear characteristics of the diode were included in the simulations: the junction capacitance, the diffusion capacitance, and the DC current/voltage relation. Bifurcation analysis and Lyapunov exponents were used to predict the limit cycles, their stability, and regions of chaos throughout the 2-dim parameter space of the driving source's frequency and amplitude. Circuit measurements were made in which the function generator's frequency and amplitude were varied so as to collect data in the same 2-dim parameter space as the simulations. The results are shown in the regime maps, Figs.\ \ref{sigma-0.01-regime map} and \ref{1N4001-regime map}. 

These two regime maps show significant agreement of simulation and circuit measurement over the 2-dim parameter space. The basic structure of the maps is the same except for the sharp dip of the $1^{st}$ PD border at twice the resonant frequency. Except for this dip, both maps show that for low source amplitude the main LC (frequency same as the source) covers the entire frequency range, down to the lowest investigated frequency (about 0.2). The bottom panels of Fig.\ \ref{time series compare} show the time-series for simulation and measurement of this LC for low source amplitude at $\omega=1$. At low source amplitude the capacitance is essentially constant: the sum of the explicit capacitor, the stray capacitance, and the diode's junction capacitance. As expected for a constant capacitance with source at resonance, the capacitor-diode voltage amplitude (orange) is much larger than the source (blue) and is nearly sinusoidal. And, varying the source frequency at low source amplitude produces the resonance peak expected for an underdamped driven RLC circuit, as seen for the 25 mV source in Fig.\ \ref{circuit-resonance}. 

A region of LPs of the main LC is apparent in both maps beginning at low source amplitude near resonance ($\omega=1$) and extending to lower frequency and higher source amplitude. These LPs are due to the hysteresis of the $\omega$-continuation seen in Fig.\ \ref{w-contin-sig.01}. As the source amplitude increases, the resonance peak distorts, creating hysteresis. This distortion is apparent in Fig.\ \ref{circuit-resonance} where the left side of the peak for the larger source amplitude is getting very steep. 

The regime maps in Figs.\ \ref{sigma-0.01-regime map} and \ref{1N4001-regime map} show similar period-doubling cascades to chaos. Time series related to the period-doublings are seen in Fig.\ \ref{time series compare} where the transition from the bottom panel to the middle panel corresponds to crossing the $1^{st}$ PD line in the regime maps as source voltage increases at $\omega=1$. The resulting period-2 LC for simulation and measurement have the same patterns. Transition from the middle panel to the top panel corresponds to crossing the $2^{nd}$ PD line which results in period-4 LC. Again, agreement on the LC shape is good between the left and right panels. Chaos inside the period-doubling cascades of the maps is predicted by the regions of positive maximal LE in Fig.\ \ref{LE}. Locations of PD lines in Fig.\ \ref{sigma-0.01-regime map} agree with the locations indicated on Fig.\ \ref{LE}. The phase plots for simulation and measurement in Figs.\ \ref{sim-phase-plot} and \ref{circuit-phase-plots}(b) look quite similar and are also consistent with chaos inside the period doubling cascades. 

The agreement discussed above between simulations and circuit measurements of the dynamical regimes suggest that the standard diode model used in SPICE and used here is robust. It was necessary to take all three nonlinear characteristics of the diode into account in order to achieve this level of agreement. The one major difference between simulation and circuit measurement in Figs.\ \ref{sigma-0.01-regime map} and \ref{1N4001-regime map} is the sharp dip of the $1^{st}$ PD bifurcation at $\omega=2$ seen in simulation but not in measurement. This same discrepancy is apparent in earlier comparisons of simulation and measurement made by Carrol and Pecora. \cite{Carroll2002} They used measurements from different diodes to make regime maps showing the onset of the $1^{st}$ PD, then compared the maps to calculated $1^{st}$ PD regime maps. Only some of the diode circuit measurements showed the PD dip at $\omega =2$, whereas nearly all the simulations showed the dip. Their result is consistent with the finding here that the measurements in Fig.\ \ref{1N4001-regime map} seem to show an upside-down shoulder at $\omega=2$, but not a dip, whereas the simulation in Fig.\ \ref{sigma-0.01-regime map} has the dip. This discrepancy of the existence of the PD dip was not a concern in the earlier work, and is an interesting observation here to be investigated in future work. 

Other early works focused on identifying the nonlinear property of the diode responsible for the period-doubling cascades to chaos. Accounting for both the nonlinear junction capacitance and diffusion capacitance predicted observed bifurcations. \cite{Brorson1983,Azzouz1983} Diffusion capacitance can account for the diode's history-dependent reverse recovery time, thereby predicting period-doubling cascades to chaos. \cite{Moraes2003} Later, a mathematically simpler model obtained from the junction capacitance alone by using a cubic approximation of Eq.\ \eqref{Q_M<1} was used to predict nonlinear phenomena seen in the varactor loaded SRRs. \cite{wang2008,lazarides2011} 

Recently, simulations inspired by applications of nonlinear SRRs were done, producing 2-dim regime maps using the maximal LE which predict period-doubling, chaos, hysteresis, and multistability. \cite{leutcho2023} Those simulations take into account the varactor diode's junction capacitance by using the cubic approximation mentioned above. There were no experimental results for comparison. 

The goal of the work presented here is to evaluate just how well diode models can predict the measured dynamical behavior over a large 2-dim parameter space. The good agreement of prediction and measurement presented here required all three nonlinear characteristics of the diode. Equation \eqref{Q_M<1} was used instead of its cubic approximation so as to not limit the range of parameter space. 

Accurate prediction of the behavior of a varactor loaded SRR should be beneficial for designing their use for application in metamaterials. The semiconductor diode model used here is applicable to the varactor diodes in SRRs. Only the values of the model parameters differ. The model's good predictive ability demonstrated here encourages application of the bifurcation methods and tools used here to the varactor loaded SRRs. 
 
\section{Conclusion}
The diode resonator is a well-known circuit demonstrating nonlinear behavior including period-doubling cascades to chaos. Earlier works have shown comparisons of simulated and measured bifurcation diagrams, however comparisons of dynamical regime maps are far less common. This work adds to the long story of analysis of the diode resonator by using bifurcation analysis tools to generate more detailed regime maps for comparison with circuit measurements. The standard diode model used in the SPICE circuit simulator gave good results when used in the bifurcation tools and Lyapunov exponent calculations. The location of period-doubling cascades, hysteresis, and chaos in the 2-dim parameter space of drive frequency and amplitude is nearly the same in the simulated and measured maps. Lyapunov exponent plots are consistent with the regime maps. The bifurcation methods and tools used to generate the regime maps may be useful to the emerging applications of nonlinear SRRs. Future work will address the $1^{st}$ PD dip discrepancy in the low-amplitude/high-frequency portion of the regime map. In addition, other values of resistive loss (R in circuit, parameter $\sigma$ in model) will be investigated. 

% If you have acknowledgments, this puts in the proper section head.
\begin{acknowledgments}
This work was supported by a University of North Carolina Greensboro Research Assignment.  
\end{acknowledgments}

\appendix*
\section{Diode Model \& Derivation of $\mathbf{v_d(q)}$}\label{derive_V(Q)}
The mathematical model for the semiconductor diode has three parts: the quasi-static (DC) current-voltage relation, a junction capacitance important for reverse bias and small forward bias voltages, and a diffusion capacitance important for forward bias. The model's parameters are provided by SPICE parameters. \cite{rect-appl-hb}

The well-known ``diode equation" gives the DC current-voltage relation. Its parameters are reverse saturation current $I_S$ and emission coefficient $N$. The mathematical model is
\begin{equation}
I_d(V_d)=I_S\left(e^{40V_d/N}-1\right)
\label{i-v_model}
\end{equation} 
where 40 comes from $e/kT=(25mV)^{-1}$ at room temperature. 

The model's parameters for the junction capacitance are the zero-bias junction capacitance $C_{J0}$, the contact potential $V_J$, the junction capacitance grading exponent $M$, and the forward bias coefficient $FC$. The junction capacitance's dependence on voltage is given by 
\begin{equation}
C_J(V_d)=C_{J0}\left(1-\frac{V_d}{V_J}\right)^{-M}.
\label{cj_model}
\end{equation}
Parameter $FC$ is in the range 0 to 1 and gives an upper limit of $FC\times V_J$ for $V_d$ in order to avoid the singularity at the forward bias value $V_d=V_J$.

The diffusion capacitance needs one additional parameter, the forward transit time $TT$. The diffusion capacitance's dependence on voltage is given by 
\begin{equation}
C_{df}(V_d)=\frac{40TT\times I_S}{N}e^{40V_d/N}.
\label{cdf_model}
\end{equation} 

We use SPICE parameter values typical for the 1N4003 rectifier diode: $I_S=70$ pA, $N=1.4$, $C_{J0}=33$ pf, $V_J=0.35$ volts, $M=0.45$, $FC=0.5$, and $TT=5\mu s$. We use the 1N4003 diode for circuit measurements because it has a larger junction capacitance than small signal diodes, thereby reducing the effects of stray capacitance. We measured the diode capacitance as a function of bias voltage for comparison with the model's predictions. 
\begin{figure} [h]
\includegraphics[width=3.2 in]{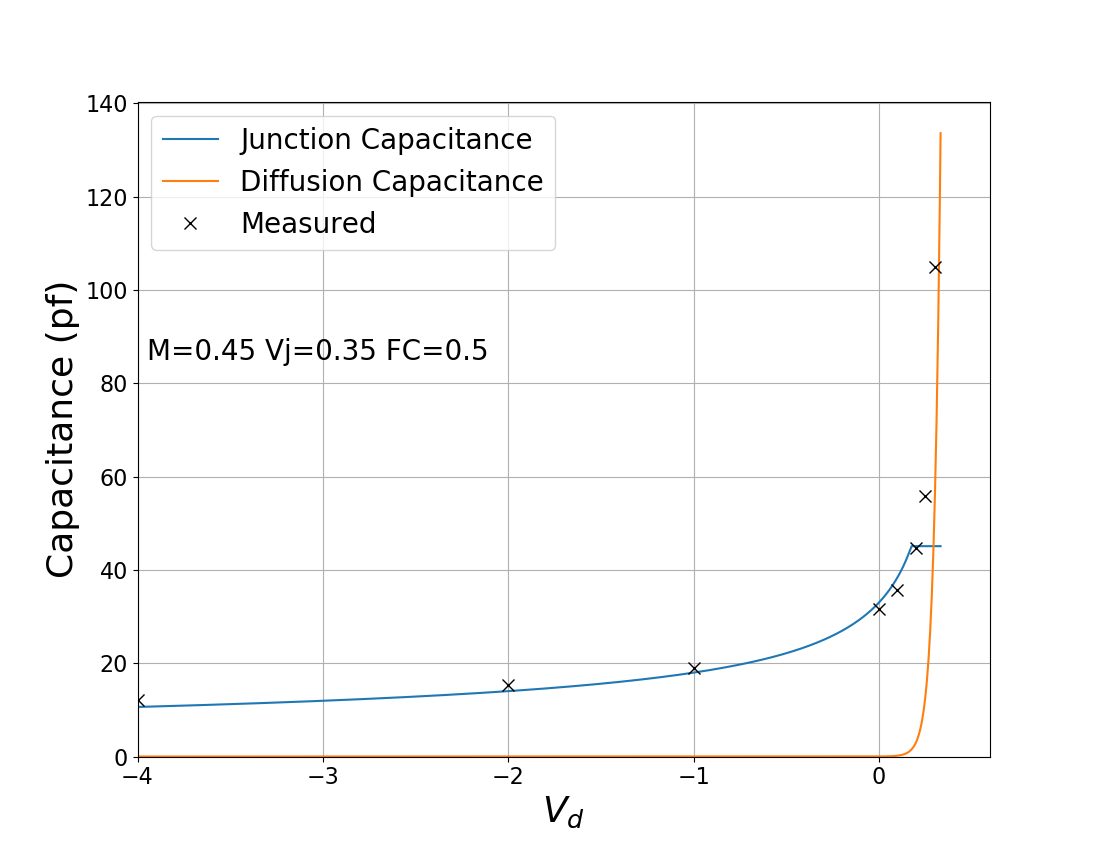}
\caption{Junction capacitance and diffusion capacitance versus bias. Measured data values (X) for 1N4003. }
\label{1N4001cap_vs_bias}  
\end{figure}
The results in Fig.\ \ref{1N4001cap_vs_bias} show our data and the models for the junction capacitance and diffusion capacitance using SPICE parameter values given above. The plot shows that junction capacitance dominates for reverse bias and that diffusion capacitance dominates for forward bias ($> 0.3$ volts). 
 
The dependence of $v_d$ on the variables $(q,i)$ is required in Eq.\ \eqref{auton-eqns}. The junction capacitance Eq.\ \eqref{cj_model} is useful because by definition 
\begin{equation}
C_J=\frac{dQ}{dV_d}=C_{J0}\frac{dq}{dv_d}. 
\end{equation}
$C_J$ is integrated to find $q(v_d)$, which is then inverted to find $v_d(q)$. \cite{wang2008,lazarides2011}  $M=1$ and $M\ne 1$ are treated separately. The case $M=1$ with the condition $q(v_d=0)=0$ results in 
\begin{equation}
q(v_d)= \ln \left(\frac{1}{1-v_d}\right).
\end{equation} 
Rearranging for voltage and using dimensionless variables finds
\begin{equation}
v_d(q)=1-e^{-q}.
\label{Q_M=1}
\end{equation}
For $M\ne 1$, integration with the condition $q(v_d=0)=0$ gives
\begin{equation}
q(v_d)=\frac{1}{1-M}\left[1-\left(1-v_d\right)^{1-M}\right].
\label{Q(V)}
\end{equation}
Rearranging for voltage finds 
\begin{equation}
v_d(q) =1-\left[1-(1-M)q\right]^{\frac{1}{1-M}}.
\label{Q_M<1}
\end{equation}

\begin{figure} [h]
\includegraphics[width=3.2 in]{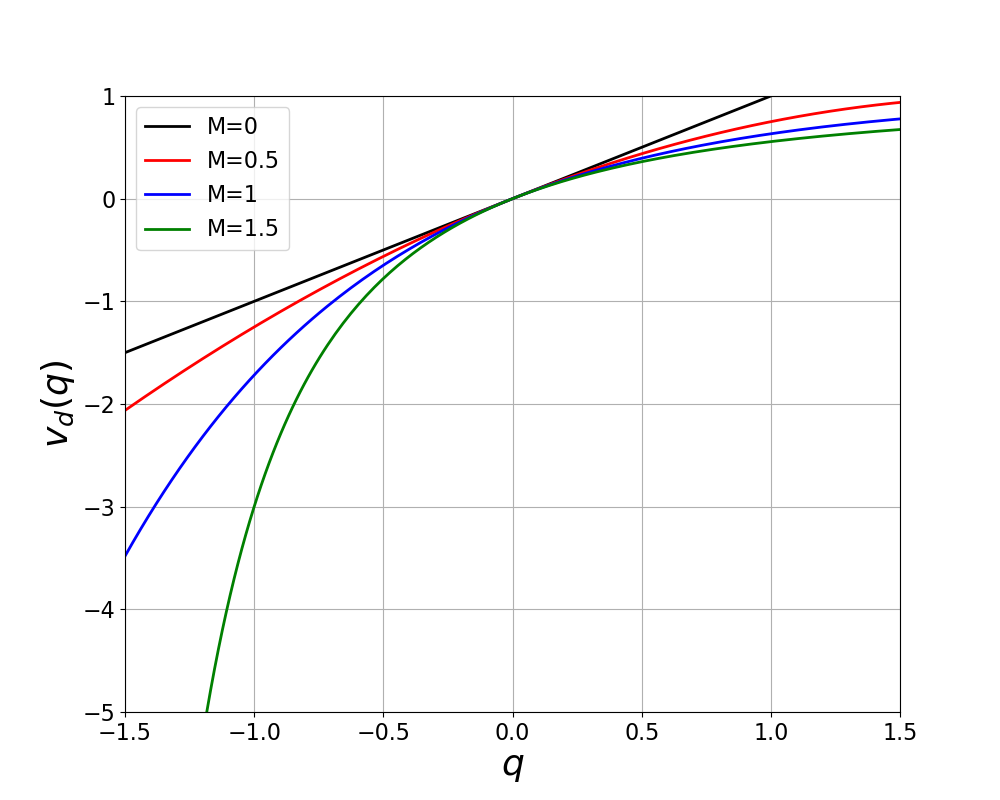}
\caption{Dimensionless diode voltage as function of junction charge for different SPICE $M$-values.}
\label{vex_compare_M-values}  
\end{figure}

Thus, the dimensionless result for $v_d(q)$ is 
\begin{equation}
v_d(q)=
\begin{cases}
1-\left(1-(1-M)q\right)^{\frac{1}{1-M}}\; & \text{for }M\ne 1\\
1-e^{-q} & \text{for }M=1
\end{cases}
\label{V(Q)}
\end{equation}
with the restriction $(1-M)q<1$. Figure \ref{vex_compare_M-values} shows plots of Eq.\ \eqref{V(Q)} for various $M$-values. 
It is important to point out that if $(1-M)q>1$ then Eq.\ \eqref{V(Q)} gives physically unrealistic complex numbers for $V_d$ since the exponent $1/(1-M)$ will in general not be an integer. These complex voltages are avoided by incorporating parameter FC. If FC is not included, then complex voltages are avoided only in carefully selected cases which give integer exponents (like $M=0.8$ giving $1/(1-M)=5$) because a negative number raised to an integral power remains real. The use of these special integral exponent cases requires caution since the imaginary parts of complex number results remain zero for $(1-M)q>1$, thereby masking the nonphysical conditions.

The unphysical singularity of the junction capacitance at $V_d=V_J$ is avoided by
letting $C_j$ remain constant for $V_d>FC\times V_J$, retaining the value $C_{J0}/(1-FC)^M$ it has at $V_d=FC\times V_J$. The $q$-value where this transition occurs is found using $v_d=FC$ in Eq.\ \eqref{Q(V)},
\begin{equation}
q_t=\frac{1-(1-FC)^{(1-M)}}{1-M}
\end{equation}
The constant junction capacitance for $v_d>FC$ makes the voltage-charge relationship linear. Therefore, Eq.\ \eqref{Q_M<1} is modified to maintain a constant slope for $q>q_t$
\begin{equation}
v_d(q)=
\begin{cases}
1-\left(1-(1-M)q\right)^{\frac{1}{1-M}}\; & \text{for }q < q_t\\
FC+(1-FC)^M(q-q_t) & \text{for }q> q_t
\end{cases}
\label{q-cases}
\end{equation}

Equation \eqref{Q-eqn} shows that $\frac{dv_d(q)}{dq}$ is required. Using Eq.\ \eqref{q-cases} 
\begin{equation}
\frac{dv_d(q)}{dq}=
\begin{cases}
\left(1-(1-M)q\right)^{\frac{M}{1-M}}\; & \text{for }q < q_t\\
(1-FC)^M & \text{for }q> q_t
\end{cases}
\end{equation}

The dimensionless DC current-voltage relation is
\begin{equation}
i_D(q)=i_S\left(e^{40V_jv_d(q)/N}-1\right)
\label{diode_current_dimless}
\end{equation}
where Eq.\ \eqref{dimensionless} is used to obtain $i_S$, the diode's dimensionless reverse saturation current. 

% Create the reference section using BibTeX:
\bibliography{nsrr-ms}

\end{document}